\documentclass[%
aip,
amsmath,amssymb,
reprint
]{revtex4-1}

\usepackage{amsthm,amsmath,color,mathptmx,times,subfiles}
\usepackage{graphicx,epstopdf,subfig}
\usepackage[x11names]{xcolor}
\usepackage{soul}
\setstcolor{blue}
\begin{document}
\newcommand{\ie}{\emph{i.e.}}
\newcommand{\edit}[1]{\textcolor{black}{#1}}
\newcommand{\guga}[1]{\textcolor{orange}{#1}}

\title{Sandpile cascades on oscillator networks: the BTW model meets Kuramoto}
\author{Guram Mikaberidze}
\email{mikaberidze@ucdavis.edu}
\affiliation{Department of Mathematics, University  of  California, Davis, CA, 95616, USA}
\author{Raissa M. D'Souza} 
\affiliation{University  of  California, Davis, CA, 95616, USA}
\affiliation{Santa Fe Institute, Santa Fe, NM, 87501, USA}

\date{\today}

\begin{abstract}
\edit{Cascading failures abound in complex systems and the BTW sandpile model provides a theoretical underpinning for their analysis. Yet, it does not account for the possibility of nodes having oscillatory dynamics such as in power grids and brain networks}. \edit{Here we consider a network of Kuramoto oscillators upon which the BTW model is unfolding, enabling us to study how the feedback between the oscillatory and cascading dynamics can lead to new emergent behaviors. We assume that the more out-of-sync a node is with its neighbors the more vulnerable it is and lower its load-carrying capacity accordingly. And when a node topples and sheds load, its oscillatory phase is reset at random}. \edit{This leads to novel cyclic behavior at an emergent, long timescale. The system spends the bulk of its time in a synchronized state where load builds up with minimal cascades. Yet, eventually the system reaches a tipping point where a large cascade triggers a ``cascade of larger cascades,” which can be classified as a Dragon King event. The system then undergoes a short transient back to the synchronous, build-up phase}. The coupling between capacity and synchronization gives rise to endogenous cascade seeds in addition to the standard exogenous ones, and we show their respective roles. We establish the phenomena from numerical studies and develop the accompanying mean-field theory to locate the tipping point, calculate the load in the system, determine the frequency of the long-time oscillations and find the distribution of cascade sizes during the build-up phase.
\end{abstract}

\maketitle

\begin{quotation}
Cascading failures and synchronization dynamics are key collective behaviors studied in complex networks \cite{BTW1987,goh2003SFsandpile,turalska2019,kuramoto1975,kuramoto1984,rodrigues2016,acebron2005kuramoto,mirollo1990synchronization,d2019explosive,newman2018networks,barrat2008dynamical,dsouza2017curtailing,zheng2017multiSP,dsouza2012suppressing,liang2010influences,dobson2007complex,carreras2004evidence,dobson2001initial,fontenele2019criticality,bellay2015irregular,shew2015adaptation,friedman2012universal,rubinov2011neurobiologically,chialvo2010emergent,beggs2003neuronal,dorfler2014synchronization,schmietendorf2014self,dorfler2012synchronization,frasca2011analysis,filatrella2008analysis,montbrio2018kuramoto,qin2019partial,menara2019framework,qin2020mediated}.  Cascading failures are often studied using ``sandpile" models \cite{BTW1987,sornette2006critical} where load arrives on nodes with limited capacity that topple once they exceed capacity and shed load to neighbors who may, in turn, topple, leading to power-law distributions in the number of topplings. Such models have been applied to a broad range of systems including power grids\cite{dsouza2017curtailing,zheng2017multiSP,dsouza2012suppressing,liang2010influences,dobson2007complex,carreras2004evidence,dobson2001initial} and brain networks \cite{fontenele2019criticality,bellay2015irregular,shew2015adaptation,friedman2012universal,rubinov2011neurobiologically,chialvo2010emergent,beggs2003neuronal} although sandpile dynamics do not account for the oscillator behavior of the nodes. In contrast, synchronization dynamics are often modeled using the celebrated Kuramoto model \cite{kuramoto1975,kuramoto1984,rodrigues2016} which, in turn, has been applied to a broad range of systems, including power grids \cite{dorfler2014synchronization,schmietendorf2014self,dorfler2012synchronization,frasca2011analysis,filatrella2008analysis} and brain networks \cite{qin2020mediated,qin2019partial,menara2019framework,montbrio2018kuramoto} although it does not account for load arrival and redistribution. To understand how cascading load dynamics interact with oscillatory nodal dynamics, we study the Bak-Tang-Weisenfeld sandpile model on a network of Kuramoto oscillators. Thus each node has two degrees of freedom: its load capacity and its phase. A natural choice of coupling is to lower the load capacity for nodes that are out-of-sync with their neighbors (to capture their enhanced vulnerability) and to randomize the phase of an oscillator after it topples (since it needs to resynchronize with the rest of the network). We show how this leads to emergent long-time oscillations not present in either model alone including ``Dragon King" events. Our study shows that emergent new phenomena can arise when distinct dynamical models are coupled together and paves the way towards understanding the interplay of cascading loads and the synchronization of oscillators.
\end{quotation}

\section{Introduction}
Cascading failures are a hallmark of complex systems, and the paradigm of self-organized criticality (SOC) provides a theoretical underpinning for studying such phenomena. A prototypical model is the Bak-Tang-Wiesenfeld sandpile model~\cite{BTW1987}  where each node of a network has a capacity to hold load above which it becomes unstable. Unstable nodes topple and shed their load to their neighbors, who may in turn topple and cause even further topplings, sometimes causing global cascades. This local redistribution of load during a failure leads to a cascade of size $s$, where $s$ follows a power-law distribution $P(s) \sim s^{-\gamma}$. The BTW sandpile model and the SOC framework have been used to study cascading failures in systems as diverse as forest fires \cite{malamud1998forest,turcotte1999self,sinha2000forest}, earthquakes \cite{turcotte1999self}, landslides \cite{turcotte1999self, hergarten2003landslides}, power grids \cite{dobson2001initial, liang2010influences, carreras2004evidence, dobson2007complex, dsouza2012suppressing, dsouza2017curtailing, zheng2017multiSP} and brain networks \cite{beggs2003neuronal, chialvo2010emergent, rubinov2011neurobiologically, friedman2012universal,bellay2015irregular, shew2015adaptation, fontenele2019criticality} (although, criticality in the brain remains a controversial topic, see Refs. \onlinecite{bonachela2010self, gross2014brain_criticality, priesemann2014spike, wilting2018inferring, yaghoubi2018neuronal, priesemann2018can, fosque2021evidence} for the discussion of the extent to which SOC applies to brain dynamics), and many more. Despite the success and widespread use of the BTW sandpile model, it neglects a fundamental aspect of power grids and brain networks, namely that the nodes in the system are oscillators with phase (and sometimes amplitude) degrees of freedom.  How the phase dynamics interact with the sandpile dynamics is a question of theoretical and practical importance that is largely unexplored.

Synchronization is another collective behavior often observed in complex systems when a collection of elements all follow the same dynamical trajectory. The celebrated Kuramoto model \cite{kuramoto1975,kuramoto1984} laid the foundation for understanding synchronization phenomenon in networks of oscillators. It has been used to model various real-world systems including power grids \cite{filatrella2008analysis, frasca2011analysis, dorfler2012synchronization, dorfler2014synchronization, schmietendorf2014self} and neuronal activity \cite{montbrio2018kuramoto, qin2019partial, menara2019framework, qin2020mediated}. Avalanches of desynchronization on networks of oscillators have attracted increasing interest in recent years. There are two general classes of oscillator cascades: phase dynamics of oscillators with complex patterns (mainly desynchronization bursts and spontaneous synchrony) \cite{poil2008avalanche, gireesh2008neuronal, yang2012maximal, coutinho2013kuramoto, maistrenko2014cascades, odor2018heterogeneity, munoz2018landau, odor2020power, liang2020hopf, gilpin2021desynchronization, munoz2021hybrid} and threshold oscillators with more-or-less uniform driving \cite{olami1992OFCmodel, cowie1993statistical, miltenberger1993fault, sornette1994statistical, corral1995self, cowie1995multifractal, chabanol1997analysis, piedrahita2013modeling, wray2014cascades, wang2016growth}. The cascades are in terms of either synchronization/desynchronization patterns, or simultaneous firing in the case of threshold dynamics, so, none of these models account for separate degrees of freedom for load and for phase.

Here we show how the interplay between the oscillatory and cascading degrees of freedom can give rise to new emergent behaviors. We couple the quintessential model of cascading failures with that of synchronization to study the BTW model on a network of Kuramoto oscillators and introduce a model that couples oscillating and load-bearing degrees of freedom. We propose a natural coupling between the models inspired by loose analogy to electric power grids. To capture the fact that generators that are out of sync with the grid are more vulnerable to a failure, we penalize nodes out-of-phase with their neighbors by lowering their capacity to hold load proportional to a parameter $\alpha$ called the desynchronization penalty. (This provides feedback from the Kuramoto dynamics to the BTW model.) After a node topples, its phase resets randomly, which mimics that generators in the electric grid need to resynchronize with the network after restarting. (This provides feedback from the BTW model to the Kuramoto.) 

Numerical simulations reveal that this coupled BTW-Kuramoto model dynamics leads to a long-time oscillatory behavior dominated by a desirable phase where oscillators are fully synchronized, and the total load builds up, while cascades are mostly avoided. However, ultimately the system reaches a tipping point where a self-amplifying mechanism kicks in and, in contrast to the BTW model, a large cascade triggers an even larger cascade, leading to a ``cascade of cascades," which can be classified as a Dragon King event \cite{sornette2009dragon_kings,sornette2012dragon}. This causes a system-wide discharge of load, reminiscent of running sandpiles\cite{hwa1992running_sp}. Then the system has a short transient dynamic that restores complete synchrony and returns the system to the build-up phase. The period of the long-term oscillations is at an emergent timescale that introduces a long-range order in the system and predictability to the Dragon King events. Two different cascade sources control such behavior: exogenous seeds that originate from the arrival of new load and endogenous seeds that originate from nodes that fail due to lowered capacity. Analytic treatment based on mean-field theory and branching processes allows us to approximate the location of the tipping point, the frequency of the long-term oscillations, and uncover a non-trivial dependence of the total load on the network degree. We also present a semi-analytic prediction of cascade size distribution in the build-up phase and discuss the system behavior in the thermodynamic limit.

Note that massive events are often referred to as being Dragon Kings or being black swans. Dragon King (DK) events are massive events which occur significantly more often than what extrapolation from smaller events would suggest. Often, Dragon Kings are related to tipping points, phase transitions, or bifurcations and are generated by a self-amplifying, nonlinear, endogenous mechanism that is different from the mechanism behind smaller events \cite{sornette2009dragon_kings, sornette2012dragon}. DKs are often easy to predict and they can be contrasted from black swans which are tail events from a power law distribution, such as in SOC. A black swan  is a metaphor that describes a rare event that comes as a surprise, has a major effect, but is ultimately generated by the same mechanism as the common events (SOC in the BTW model) and is therefore hard to predict. Both, DKs and black swans are massive events, but the DKs occur with much higher probability and are much more predictable.

This manuscript is organized in the following manner. Section \ref{sec_background} introduces the basics of the BTW sandpile and Kuramoto models. Section \ref{sec_model} presents the detailed formulation of the coupled BTW-Kuramoto model and the corresponding symmetry transformations. Section \ref{sec_rules} establishes the main phenomena observed through numerical simulations. Section \ref{sec_analytic} presents the analytical calculations of some essential aspects of the model. Finally, Section \ref{sec_discussion} contains discussion and conclusions.

\section{Background} \label{sec_background}

\subsection{BTW Sandpile model}

The Bak-Tang-Wiesenfeld (BTW) Sandpile model \cite{BTW1987} is a discrete dynamical system originally defined on a 2D square lattice of nodes and later generalized to many other lattice and network topologies. Each node $i$ in the system has some capacity $c_i$ to accommodate the load, and at any given instant, it holds some integer amount of load $s_i$. At each discrete time step, a grain of sand (i.e., load) falls on a randomly selected node $i$. If the load now exceeds the capacity ($s_i > c_i$), the node is unstable, and it topples, shedding $c_i+1$ units of load evenly to its neighbors. (In the original BTW model, the nodes have capacity $3$, so each of the $4$ neighbors receives exactly $1$ unit of load after toppling.) Some of those neighbors may, in turn, become unstable and topple, leading to further toppling, causing an avalanche of topplings. Regardless of the extent of a cascade, it is allowed to finish before the next unit of load is added to a randomly selected node. Each cascade can be measured by its ``size," i.e., the number of topplings, or by its ``area," i.e., the number of distinct nodes that toppled. For locally tree-like networks, these measures are equivalent in the thermodynamic limit of $N\rightarrow\infty$.

Load must be allowed to leave the system in some way. Otherwise, the system will eventually become supercritical, and a never-ending cascade will begin. Usually, this is achieved by either imposing open boundary conditions on the lattice or introducing a small dissipation probability $p$ for each flowing unit of load. If the system has a small amount of total load $S$, only small cascades will happen, causing little dissipation, so $S$ will increase due to load arrival. In contrast, if the total load $S$ is very large, extremely large cascades will happen, dissipating much load and reducing $S$. These two tendencies balance perfectly in the stationary state, which is achieved for some specific intermediate value of total load $S=S_c$.

The balance of load arrival and dissipation leads to a power law (i.e., ``critical") distribution of event sizes in the thermodynamic limit. For networks with a wide range of different topologies the distribution of avalanche sizes obeys $P(s) \sim s^{-\gamma}$ with $\gamma=\frac{3}{2}$. The phenomenon is dubbed ``self-organized criticality (SOC)." The largest events are commonly referred to as ``black swans" \cite{taleb2007black} since they are rare, massive events generated by the same mechanism as smaller events (i.e., SOC).

A system is considered critical if its correlation length diverges and power-law distributions appear. The relevance of criticality comes from its applicability to second-order phase transitions in thermodynamics \cite{pathria} and the potential to pinpoint the endings of renormalization group flows in the Standard Model of particle physics and field theories in general \cite{amit2005field,poland2019conformalBootstrap}. The divergence of the correlation length implies scaling symmetry which places strong restrictions on the system and the types of possible behavior characterized by the critical exponents and the slopes in the power-law distributions. Based on these scaling properties, various systems are grouped into ``universality classes". Achieving a critical state in thermodynamics usually requires tuning of a control parameter. But criticality in the BTW model is achieved without any tuning, and thus the name ``self-organized criticality". One can think of such a system as having a critical point as an attractor.

It should be noted that there are many variants of sandpile models such as BTW \cite{BTW1987} and Manna \cite{manna1991} sandpiles, Oslo ricepiles \cite{frette1996oslo}, etc. Here we use the BTW sandpile model, with some minor modifications.

\subsection{Kuramoto model}
The Kuramoto model \cite{kuramoto1975,kuramoto1984} is a dynamical system that was originally proposed to study the synchronization of oscillator nodes connected to one another via a complete graph. It has since been generalized to many other network topologies. Each oscillator node has a phase $\phi_i$ which evolves in time according to the equation
\begin{equation}
	\dot\phi_i(t) = \omega_i + k \sum_{j\in \mathcal{N} _i} \sin(\phi_j(t)-\phi_i(t)).
\end{equation}
Here $\omega_i$ is the natural frequency of the oscillator $i$, the summation is over the set of its neighbor oscillators $\mathcal{N} _i$, and $k$ is a parameter called the coupling constant. With a lack of coupling ($k=0$), each oscillator moves at its natural frequency. By setting $k>0$, attraction is introduced between neighbors. For a low value of coupling and a broad spectrum of frequencies $\{\omega_i\}$, the oscillators move incoherently, while for a high coupling $k$ and narrow frequency spectrum, the oscillators synchronize with each other.

It is useful to define the Kuramoto order parameter $r$ of the system of oscillators, which is a measure of overall synchrony:
\begin{equation}\label{r}
	r = \frac{1}{N} \left|\sum_{j=1}^{N}e^{i\phi_j(t)}\right|.
\end{equation}
The summation here is over all the oscillators, and $N$ is their number. When $r$ is close to 0 (or 1), the system is incoherent (or fully synchronized). In the thermodynamic limit ($N\rightarrow\infty$), the stationary value of the synchronization order parameter, Eq. \eqref{r}, displays a phase transition with $r(k)=0$ for $k<k_c$ and $r(k)>0$ for $k>k_c$. The critical value $k_c$ and the properties of the phase transition depend on the distribution of natural frequencies and the network topology \cite{rodrigues2016,d2019explosive}. 

We also define the local synchronization of node $i$:
\begin{equation}\label{r_i}
	r_i = \frac{1}{1+|\mathcal N_i|} \left|\sum_{j\in \mathcal N_i \cup \{i\}}e^{i\phi_j(t)}\right|
\end{equation}
where $\mathcal{N} _i$ is the set of neighbors of node $i$. This parameter $r_i$ is an analog of $r$ for the immediate neighborhood of node $i$.

\section{The coupled BTW-Kuramoto model (BTW-KM)}\label{sec_model}

\begin{figure}\centering
	\includegraphics[width=1\linewidth]{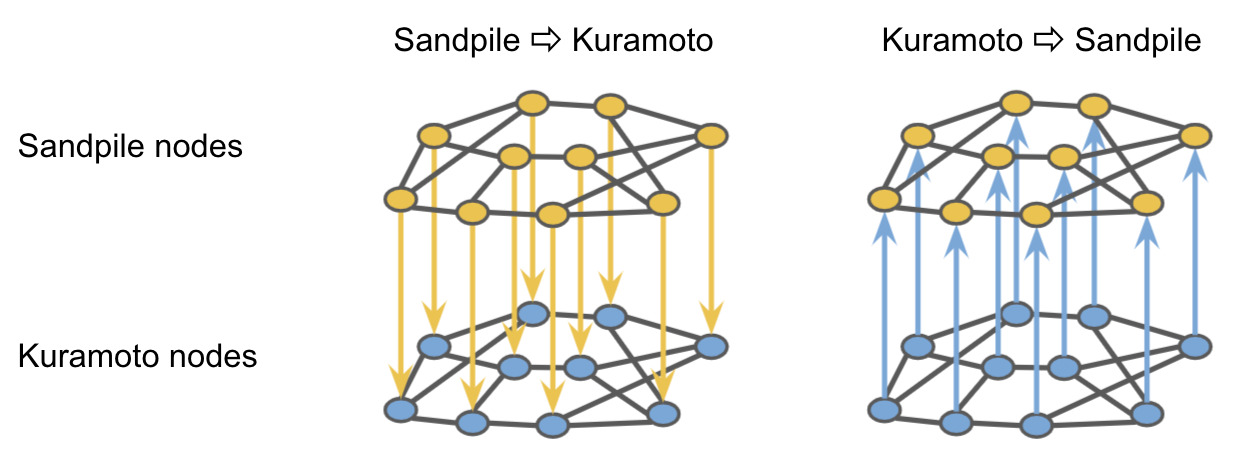}
	\caption{Visualization of  BTW sandpile and Kuramoto model interactions as a multiplex network.}
	\label{2-layer net}
\end{figure}

The coupled BTW-Kuramoto model can be visualized as a two-layer multiplex network with each layer denoting one of the subsystems and the inter-layer edges representing couplings between subsystems (Fig. \ref{2-layer net}). If we consider this as one system, each node $i$ in the network will have two degrees of freedom: an oscillator phase $\phi_i$ and a load $s_i$. This contrasts with the approach taken by Corral et al. \cite{corral1995self} where the authors consider a lattice of pulse-coupled Olami-Feder-Christiansen \cite{olami1992OFCmodel} oscillators with non-linear driving similar to the Mirollo-Strogatz model \cite{corral1995self}. This produces cascades of energy (i.e., load) on a lattice of oscillators, but in contrast from BTW-KM, the oscillator phases can be uniquely determined from energies, and therefore each node has effectively one degree of freedom. 

One fundamental issue with coupling the BTW and the Kuramoto models is that the KM is a continuous dynamical system, while BTW is a discrete-time process. To resolve this, we assume a constant rate of load arrival $\frac{1}{\Delta T}$ (one unit of load arrives per each time interval $\Delta T$). We allow the KM layer to evolve for this fixed time interval $\Delta T$ before the arrival of each unit of external load in the BTW layer. We maintain an essential feature of the BTW model: cascading dynamics are infinitely faster than the load arrival rate. So, a cascade unfolds essentially instantaneously once a unit of load arrives. The Kuramoto coupling parameter sets the synchronization rate of the oscillators to be $\frac{1}{k}$. The stronger the coupling, the faster the synchronization. Thus both $\Delta T$ and $k$ determine the relative timescale as discussed further in Sec. \ref{symmetries}.

The timescales discussed here capture some but not all the details of real electric power grids. In real-world power grids, the load variations occur on the timescale of hours and are the slowest aspect of cascading blackouts \cite{yao2015multi}. This is similar to BTW-KM dynamics, where load arrives at rate $\frac{1}{\Delta T}$ and the total load capacity of the network is proportional to its size, so, filling it up takes time $O(N\Delta T)$. In real-world power grid systems, cascading outages requiring intervention by human dispatchers typically last for tens of minutes \cite{yao2015multi}, and synchronization typically occurs on the timescale of one second \cite{liu2012oscillation}. This relation is switched in the BTW-KM where cascades are instantaneous and the phase synchronization timescale is set by $\frac{1}{k}$ for fixed $\Delta T$. So, two of the timescale relations in BTW-KM match the power grid dynamics, and one does not. However, it should be noted that our goal is not a faithful reproduction of power grid dynamics. Instead, we aim to establish a formal way to understand the interplay between cascading loads and the synchronization of oscillators.

\subsection{Rules}\label{rules_sec}

\edit{The model takes some inspiration from electric power grids.} For a power grid to operate properly, the generator phases need to be synchronized \cite{motter2013spontaneous,manson2015case}. This means that incoherent generators raise the risk of failure. We lower the capacities of out-of-sync nodes to reflect this increased risk. \edit{When a generator does fail, it must re-establish synchronization with the rest of the network once restarting. We account for this by randomly resetting the phase of each toppled node at the end of a cascade. The rules of the dynamics are:}

\begin{enumerate}
	\item \textbf{KM:} 
	Evolve KM for time $\Delta T$ according to the equation:
	$\dot\phi_i(t) = \omega + k \sum_{j\in \mathcal{N} _i} \sin(\phi_j(t)-\phi_i(t))$.
	
	\item \textbf{KM$\rightarrow$BTW:} 
	Calculate the local synchronization $r_i$ and corresponding capacities $c_i = c^0 - \alpha (l-r_{i}).$
	
	\item  \textbf{BTW:} 
	\edit{Choose a node uniformly at random and add a unit of load to it}. If any node is overcapacity, topple it, and perform the subsequent BTW cascading dynamics.
	
	\item \textbf{KM$\leftarrow$BTW:}
	Reset the phase of each toppled node $\{i_k\}$ randomly:
	$\phi_{i_k} = rand(-\pi, \pi)$.
	
	\item repeat.
\end{enumerate}

Note that the phases of the toppled nodes are reset only once the cascade finishes, and the capacities $c_i$ are calculated only before the cascade starts. So throughout the cascade the capacities do not change. Alternatively, one could execute rules 2-4 synchronously, randomizing phases and recalculating capacities while the cascade is unfolding. This yields qualitatively similar behavior that we comment on in Sec. \ref{pcdk_section}. Note also, that the frequencies of all nodes $\omega$ are set equal to each-other for simplicity.

The variables in the second step are defined as follows. First, $c^0$ is the unperturbed BTW capacity of a node. The parameter $\alpha$ is the desynchronization penalty determining how strongly the KM synchronization $r_i$ affects BTW capacities. The $(l-r_i)$ term is a measure of local desynchronization where $l$ is a constant close to 1. For a finite system, synchronization is almost never perfect, so, in the fully synchronized state $r_i$ is close to but not equal to $1$. \edit{Notice that only the integer parts of the capacities affect the cascade dynamics. Therefore if l=1, any small fluctuation away from full synchronization would result in the capacities being decremented. To accommodate for up to 10\% fluctuations of the global synchronization we set l=0.9}. This does mean that the capacities $c_i$ can increase above $c_0$ when $r_i>0.9$, however, for $0<\alpha<10$, $c_i$ will effectively be $c_0$ even when fully synchronized since only the integer part matters. Details of parameter choices will be discussed in Sec. \ref{params_and_simul_details}.

Since the capacities of nodes can vary, the BTW sandpile dynamics must be slightly modified. It will no longer necessarily be true that a toppling node has $d$ units of load where $d$ is the node degree. So, it will not always be possible to distribute the shedding load to the neighbors evenly. Therefore during each toppling, the node sheds its total load to its neighbors as evenly as possible, choosing the ones that will receive more, less, or no load at random. (See Manna model \cite{manna1991} for a similar treatment of the same issue.)

\begin{figure}[t]\centering
	\subfloat[]{\label{sweep:a}\includegraphics[width=.5\linewidth]{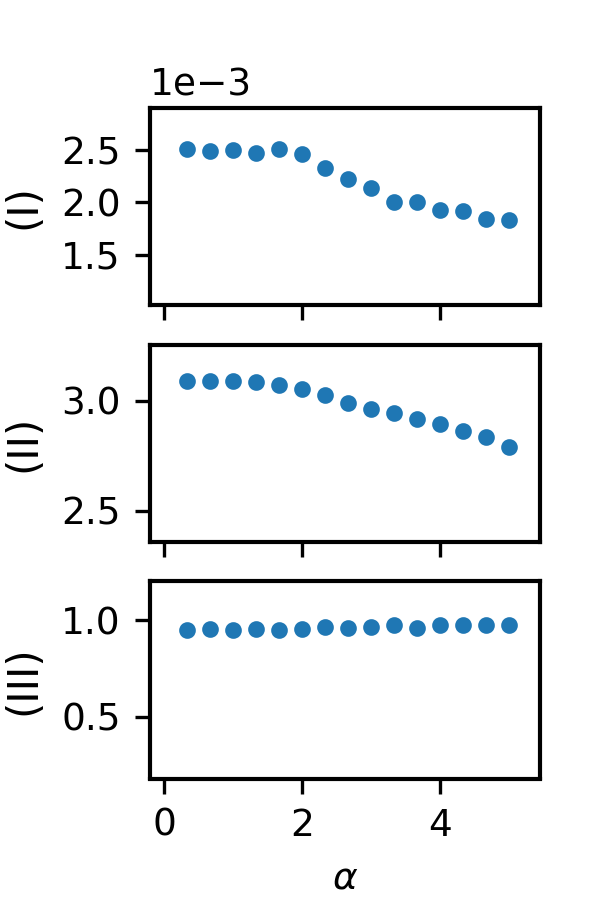}}\hfill 
	\subfloat[]{\label{sweep:b}\includegraphics[width=.5 \linewidth]{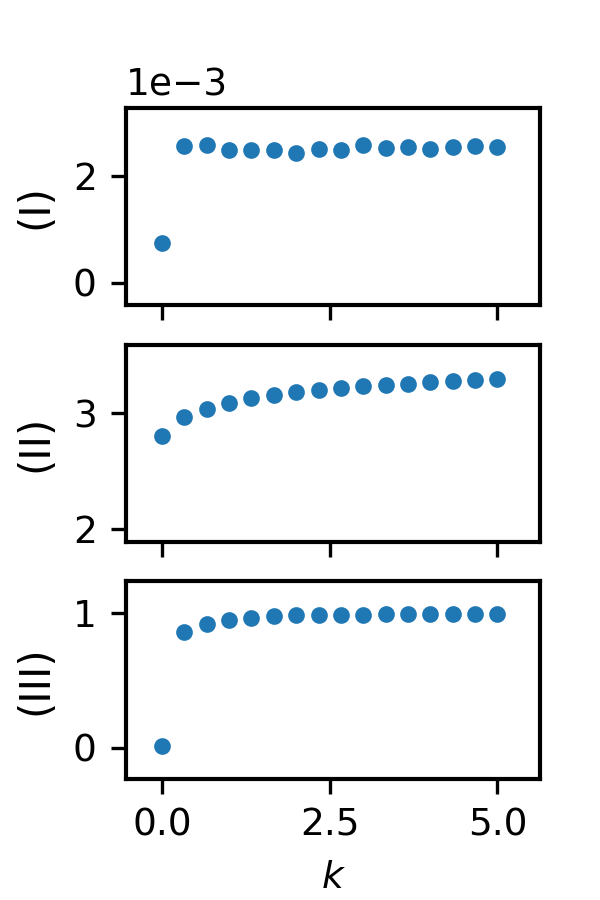}}
	\caption{Sweep of the parameter space for BTW-KM. $X$-axes: (a) desynchronization penalty $\alpha$, (b) oscillator coupling $k$. $Y$-axes: (I) probability of a large cascade that covers at least half the network, (II) average amount of load per node, (III) average global synchronization order parameter. Simulations were done on a 3-regular random network with $N=4000$ nodes, $c^0=5$, $p=0.005$, and $\Delta T=1$. Additionally, we set $k=1$ in (a), and $\alpha=1$ in (b). Error bars are smaller than the markers.}
	\label{sweeps}
\end{figure}

\subsection{Symmetries allow dimension reduction}\label{symmetries}
We can take advantage of the symmetry transformations of this dynamical system to limit consideration to a subset of the full parameter space and then generalize the results. The symmetries will also prove themselves useful in the discussion of the thermodynamic limit in Sec. \ref{sec_thermo}. The symmetry transformations are:

\begin{enumerate}
	\item Rotation of the reference frame: $\{\phi_i(t)\}\rightarrow \{\phi_i(t)+\phi_0\}$
	\item Changing the frequency: 
	$\begin{cases}
		\{\phi_i(t)\}\rightarrow\{\phi_i(t)+\omega_0 t\} \\
		\omega\rightarrow\omega-\omega_0
	\end{cases}$
	\item Reflection: 
	$\begin{cases}
		\{\phi_i(t)\}\rightarrow\{ -\phi_i(t)\}\\
		\omega\rightarrow-\omega
	\end{cases}$
	\item Time scaling: 
	$\begin{cases}
		t\rightarrow\frac{t}{\lambda}\\
		\Delta T\rightarrow\frac{\Delta T}{\lambda}\\
		k\rightarrow\lambda k
	\end{cases}$
\end{enumerate}

We also note that punishing the out-of-synch nodes by decreasing their capacity $c_i=c^0 - \alpha (l-r_i)$ is equivalent to rewarding the synchronized nodes by increasing their capacities $c_i=c^0+\alpha r_i=(c^0+l\alpha)-\alpha (l-r_i)$. So the rewarding model with unperturbed capacities $c^0$ maps to a punishing model with unperturbed capacities $(c^0+l\alpha)$.

\begin{figure}[b]\centering
	\subfloat[]{\label{N,p choice:a}\includegraphics[width=.5\linewidth]{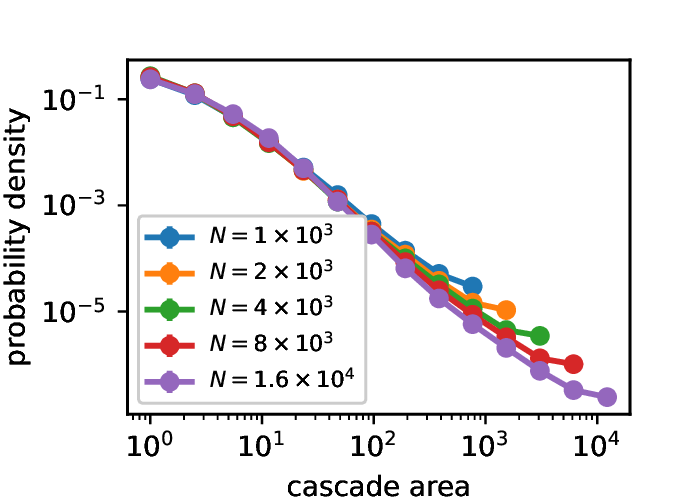}}\hfill 
	\subfloat[]{\label{N,p choice:b}\includegraphics[width=.5 \linewidth]{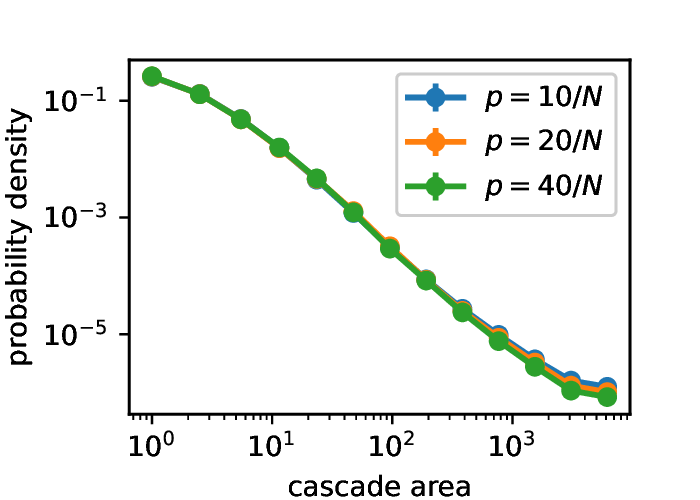}}
	\caption{Cascade area distributions for BTW-KM for (a) different network sizes $N$, and (b) different values of dissipation $p$. The data shown here are the averages over ensembles of simulations, each consisting of 30 realizations. Error bars are smaller than the markers.}
	\label{N,p choice}
\end{figure}

\section{Behavior of BTW-Kuramoto model}\label{sec_rules}
In this section, we establish the main phenomenology associated with coupled BTW-KM dynamics. We show how the system quickly synchronizes then slowly builds up load until it reaches a tipping point when a large cascade triggers a subsequent chain of even larger cascades (i.e., a cascade of cascades) that deplete the system of load and leave it desynchronized. Then the cycle begins again. We will show the endogenous, self-amplifying nature of the chain of cascades and that this chain of cascades can be classified as a ``Dragon King" (DK) event.

To focus on the emergent BTW-KM phenomenology, we keep the network structure simple and consider random 3-regular networks and various system parameters. Random 3-regular networks reasonably capture degree distribution aspects of power grids, although they have much smaller clustering coefficient and have smaller diameter (e.g., see Table S1 in Ref. \onlinecite{dsouza2012suppressing}).

\subsection{Parameter choices}\label{params_and_simul_details}
We first perform a parameter sweep to determine how the coupling between the BTW and Kuramoto models affects the system's performance over a broad range of parameters. Our objectives are to minimize large cascades and handle as much load as possible while keeping the oscillators synchronized ($r\simeq1$). Figure \ref{sweeps} shows how these parameters depend on the desynchronization penalty $\alpha$ (left column) and oscillator coupling $k$ (right column). The top row is the probability of large cascades, the second is the average amount of load in the system, and the third shows the average global synchronization. We see that increasing $\alpha$ reduces the probability of large cascades at the price of decreasing the network's total capacity. While increasing $k$ improves global synchronization and total capacity. We settle on the parameter choices $\alpha=1$ and $k=1$ for the remainder of our study unless explicitly stated otherwise.

\begin{figure}[b]\centering
	\subfloat[]{\label{S and r vs t:a}\includegraphics[width=.5\linewidth]{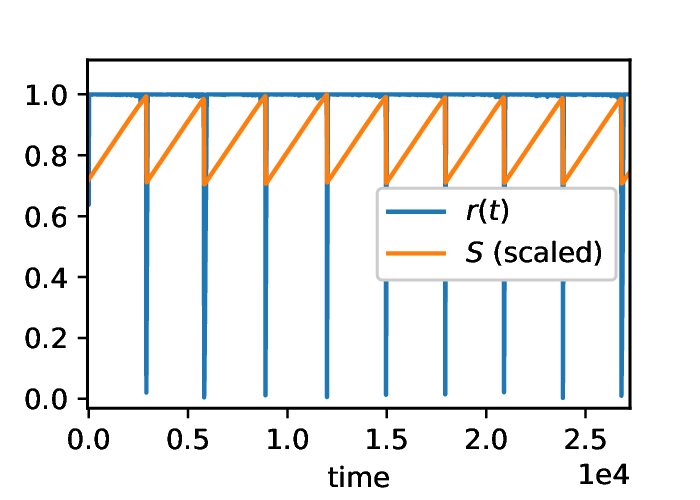}}\hfill 
	\subfloat[]{\label{S and r vs t:b}\includegraphics[width=.5 \linewidth]{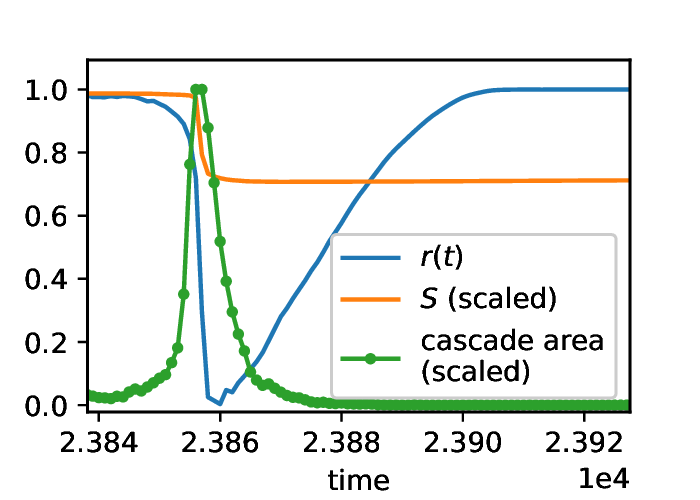}}
	\caption{Emergent long timescale oscillations in BTW-KM. (a) \edit{Time evolution of global synchronization $r$, and the total load $S$. (b) Time evolution of $r$, $S$, and cascade area on an expanded time-scale around one DK event.}}
	\label{S and r vs t}
\end{figure}

As mentioned, the BTW model has dissipation parameter $p$. For our simulations, we choose dissipation probability $p\sim20/N$ based on the rule of thumb stated explicitly in Ref. \onlinecite{dsouza2012suppressing}. Figure \ref{N,p choice:a} shows the cascade area distribution for different network sizes displaying a stable distribution shape and right tail that moves further out with increasing system size $N$. Figure \ref{N,p choice:b} shows that changing the dissipation probability $p$ does not have much effect on the cascade area distribution.

\begin{figure*}
	\centering
	\subfloat[]{\label{DKplot:a}\includegraphics[width=.33\linewidth]{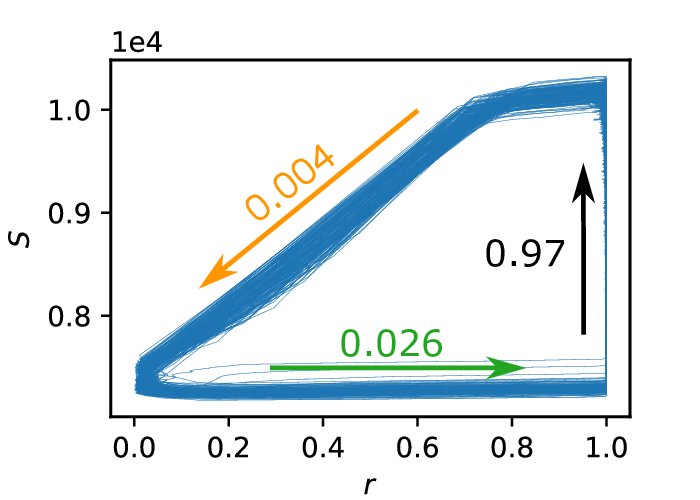}}\hfill 
	\subfloat[]{\label{DKplot:b}\includegraphics[width=.33\linewidth]{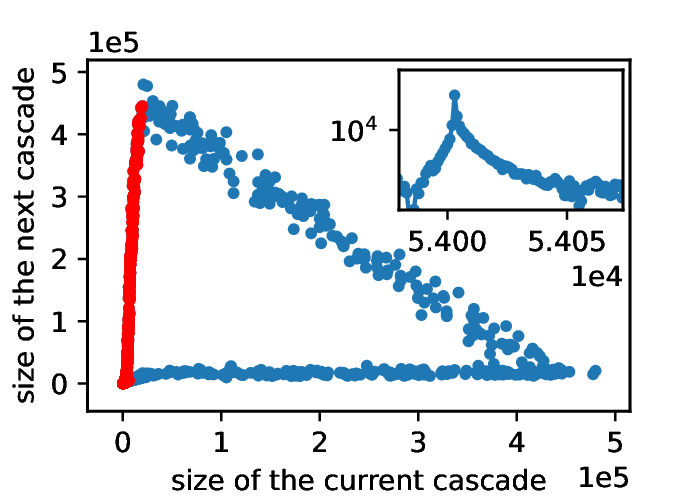}}\hfill 
	\subfloat[]{\label{DKplot:c}\includegraphics[width=.33\linewidth]{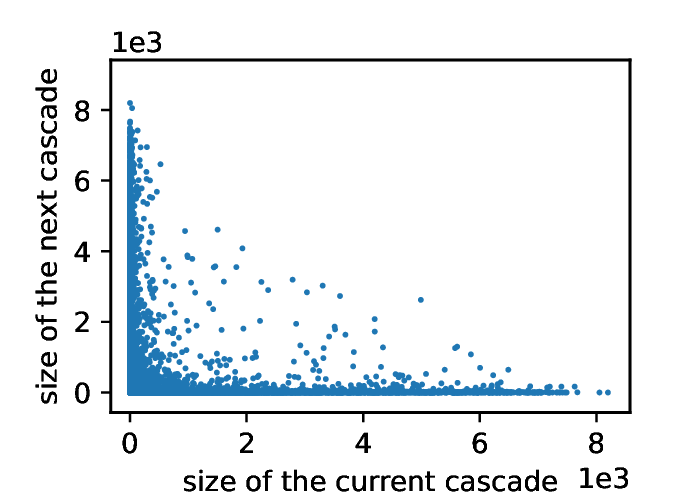}}
	\caption{(a) Total load $S$ vs. synchronization $r$. \edit{The plot shows BTW-KM cycle with its three phases: DK phase (the orange arrow); synchronization phase (green arrow); and load buildup phase (black arrow). The numbers next to each arrow indicate the fraction of time spent in that phase. (b) Evolution of cascade sizes plotted in the spirit of discrete dynamical systems. Red points correspond to pairs of events after which the synchronization order parameter is still $r>0.1$ and constitute the DK event. The inset shows time evolution of logarithm of cascade sizes around one DK event.} (c) Sizes of consecutive cascades for the BTW sandpile model.}
	\label{DKplot}
\end{figure*}

The initial conditions also need to be specified: We start with no load and random oscillator phases. Unless specified otherwise, simulations are done with 3-regular random networks of $N=8000$ nodes with the parameter $\alpha=1$. We use the Kuramoto coupling $k=1$, and $\Delta T=1$. The latter is an arbitrary choice due to the time scaling symmetry. The capacities $c^0=2$ are standard for the deterministic BTW model with degree $d=3$. And $p=0.0025$ follows from the rule of thumb \cite{dsouza2012suppressing}.


\subsection{Periodic cycles of Dragon Kings }\label{pcdk_section}
The BTW-KM dynamics has emergent periodic behavior. Figure \ref{S and r vs t} is the time-series data of the total load $S$ and synchronization $r$ obtained from a single realization. It shows a significant periodic drop in both variables. It occurs during the cascades of cascades constituting the Dragon King (DK) event. System-wide synchronization is re-established quickly, but it takes considerably longer to replenish the load that was drained during the DK event.

The emergent periodic cycles are best observed in Fig. \ref{DKplot:a} by plotting the parameters that measure different aspects of global order against each other: the KM global order parameter $r$, and the total system load $S$. Here (and in all future figures), we show combined results from ensembles of simulations. Each simulation starts with a new random regular network and new initial conditions. We see three distinct phases in this cycle: the orange arrow indicates the DK phase, green is the synchronization phase, and black is the load build-up phase. The number by the arrow is the relative fraction of the cycle spent in that phase.

Figure \ref{DKplot:b} provides insight into the dynamics during a DK event. Here we plot the sizes of consecutive cascades against each other. The points are colored red if, after both cascades, the system still has some level of synchronization ($r>0.1$). This way, we focus on the cascades that form the bulk of the DK event while excluding the last cascades when the DK phase is dying out. The red points are roughly linearly aligned with the slope $a>1$. This indicates self-amplifying feedback of cascade sizes $s(t+\Delta T)=a\ s(t)$ that causes an exponential growth in subsequent cascades
\begin{equation}\label{expo_growth}
	s(t)\propto \exp\left(\frac{\ln a}{\Delta T}t\right).
\end{equation}
The sequence of self-amplifying cascades, where each cascade causes an even larger cascade, continues until the dissipative nature of the sandpile dynamics eventually dissipates sufficient load, and the DK phase ends. Then the system returns to restoring the synchronization and accumulating load.  Figure \ref{DKplot:b} can be contrasted with the same plot for the BTW sandpile shown in Fig. \ref{DKplot:c} where there is neither a positive correlation between sizes of subsequent cascades nor DK events.

The underlying mechanism behind the self-amplifying exponential growth in Eq. \eqref{expo_growth} is the emergence of endogenous cascade seeds. In the standard BTW sandpile model, each cascade is initiated by the external addition of a unit of load (i.e., an exogenous seed). However, the BTW-KM dynamics introduces one more mechanism for initiating a cascade: a node may topple due to reduced capacity (i.e., an endogenous seed). Note that after a cascade completes, the phases of all toppled nodes are reset at random, thereby decreasing the local synchronization for these nodes and their immediate neighbors (the boundary of the cascade). The capacities of these nodes may thus be reduced, potentially turning them into endogenous seeds for the next cascade. The larger the cascade, the more endogenous seeds it can create, and multiple endogenous seeds start an even larger cascade. 

\begin{figure}[b]\centering
	\subfloat[]{\label{seed_plots:a}\includegraphics[width=.5\linewidth]{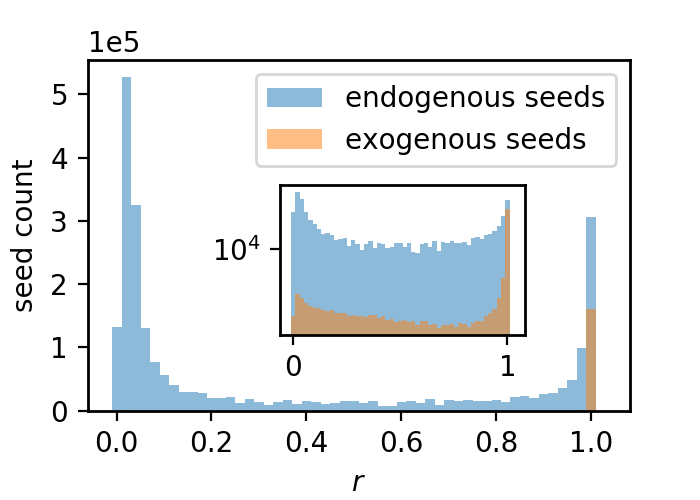}}
	\subfloat[]{\label{seed_plots:b}\includegraphics[width=.5 \linewidth]{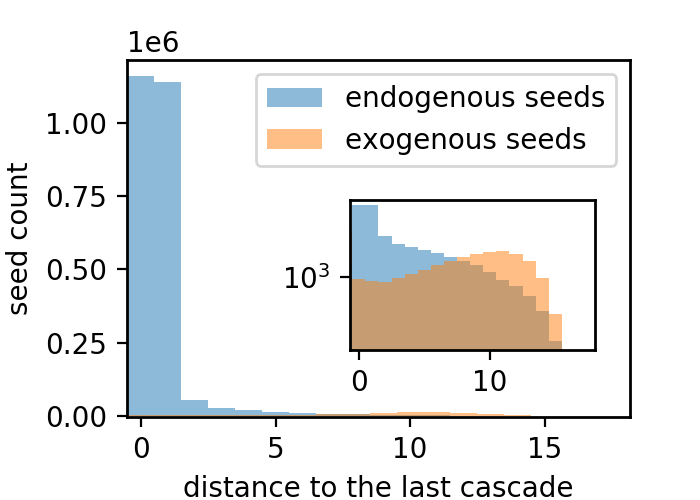}}\hfill 
	\caption{Histograms of endogenous and exogenous seeds binned by (a) synchronization order parameter $r$ at the moment of toppling (b) geodesic distance from the seed to the last cascade. The insets are the same plots with logarithmic $Y$-axis to better show the distribution of exogenous seeds.}
	\label{seed_plots}
\end{figure}

This amplification of endogenous seeds causes the chain of cascades with exponentially growing sizes leading to full system-wide cascades. This nonlinear, self-amplifying mechanism causes the most significant events in the system, turns on once reaching a tipping point where the system has a critical level of load, and differs from the standard BTW cascade mechanism. Furthermore, due to the periodicity of the BTW-KM cycle, these events arise at predictable times. These are the requirements to be classified as a Dragon King.


Figure \ref{seed_plots:a} is a histogram showing that endogenous seeds dominate DK events. We see that in states without complete synchrony ($r<1$), the number of exogenous seeds (orange) is dwarfed by the number of endogenous seeds (blue). The inset shows the same histogram on a logarithmic scale, where it can be seen that the exogenous seeds are still present in desynchronized phases of the cycle but are much fewer than endogenous ones.

To understand the locations of endogenous seeds during DK events, we look at the count of seeds as a function of the geodesic distance from the last cascade (i.e., the distance to the closest node that toppled during the last iteration). As expected, Fig. \ref{seed_plots:b} shows that the distance is most likely $0$ or   $1$, i.e., the endogenous seeds are found either within the last cascade or on its boundary. In contrast, the location of exogenous seeds seems to be independent of the last cascade: it peaks at approximately $10$, close to the average distance between two random nodes.

\begin{figure}\centering
	\includegraphics[width=.7\linewidth]{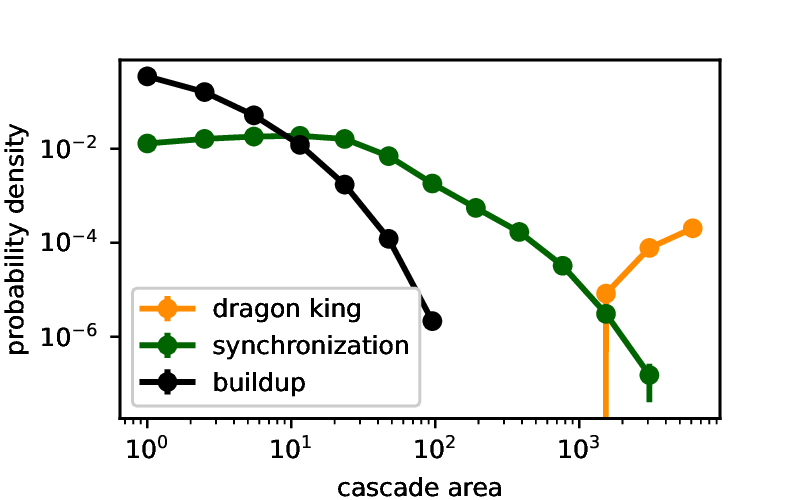}
	\caption{Cascade area distribution grouped by the phases of BTW-Kuramoto model. We decide which phase a cascade belongs to based on its location in the $(S,r)$ plane shown in Fig. \ref{DKplot:a}. We excluded the events near the crossover regions to avoid ambiguity. The error bars here and in later plots (often too small to be seen except in the legends) are the standard error of the ensemble measurements. Ensembles consist of $30$ realizations.}
	\label{areas_by_phases}
\end{figure}

Figure \ref{areas_by_phases} shows the cascade area distribution for each phase of the BTW-KM cycle separately. The long-lasting build-up phase has the most desirable cascade area distribution with the smallest average cascades. We will investigate this further in the next section. 

It should be noted that as per rules described in Sec. \ref{rules_sec}, the node capacities are kept constant during an individual cascade. In contrast, if during the cascade, the node phases are immediately randomized and their capacities are immediately recalculated upon toppling, the phenomenology of the BTW-KM remains unchanged. Particularly the emergent cycle and the self-amplified DK events are still observed. However, the chain of cascades constituting the Dragon King event unfolds from only one exogenous seed arrival which collapses the timescale of the DK phase. In contrast, the model implemented here, which updates the capacities only once before every unit of load arrives, naturally splits up the DK into a chain of cascades, allowing us to quantify the exponential growth and its underlying mechanism clearly. 

Cascade sizes during the DK phase achieve larger values than the network size (see Fig. \ref{DKplot:b}). This is because nodes can topple multiple times during one cascade reminiscent of running sandpiles\cite{hwa1992running_sp}. This implies that cascade area and cascade size will differ considerably during the DK phase (see Ref. \onlinecite{noel2013controlling} for details of the distinction between cascade size and area for the BTW model on networks). This multiple toppling behavior persists even in the thermodynamic scaling limit developed in Sec. \ref{sec_thermo}. Multiple topplings can be avoided by forcing toppled nodes to remain toppled until the end of a given cascade. In that case, it can happen that all neighbors of an unstable node have already toppled and can not accept load. Since there are no recipients, the load instead must be dissipated. This modification ensures that each node involved in a cascade topples only once, and that the area is equal to size. However, similar to the other modification with respect to when phases are reset, this change does not affect the qualitative phenomenology of BTW-KM either and the three-phase cycle is still observed.

\section{Mean-field analytic treatment}\label{sec_analytic}

Various aspects of our coupled BTW-Kuramoto model are amenable to mean-field analytic calculation. First, we must consider an unexplored aspect of the original BTW model. Namely, we calculate the BTW sandpile model's total load in random regular networks during the stationary state. Using that calculation, we can locate the tipping point of the BTW-KM and also estimate the frequency of emergent cycles. We also calculate the cascade area distribution during the long-lasting build-up phase. Finally, we discuss the behavior of the system in the thermodynamic limit.

\subsection{Total load capacity of the BTW Sandpile model} \label{mft SP}

In order to locate the tipping point and calculate the frequency of the emergent BTW-KM cycle, first, one must know the critical value of the total load $S_c$ for the original BTW sandpile model. This is because the load distribution in the coupled BTW-Kuramoto model at two key points in the emergent cycle (i.e., just before and just after the DK phase) is well approximated by pure BTW load distribution with correctly chosen capacities. We start by considering a BTW sandpile on a $d$-regular random network with fixed capacities $c$ under mean-field theory (MFT) assumptions. To find the load distribution, we consider a cascade during the stationary state, more specifically, one typical toppling during such a cascade. Being in a stationary state requires that the state just before such a toppling has the same load distribution as the state just after it.

Let us begin by fixing the capacities as per convention that $c=d-1$, where $d$ is the degree of the network. In such a system, a toppled node sends exactly $1$ unit of load to each of its neighbors. Later we will generalize the result to arbitrary capacities. Let $\rho_i$ denote the fraction of nodes that hold $i$ units of load. We will solve for $\rho_i$ in the stationary state.

We know that for the BTW model, the cascade distribution is critical. This means that, on average, each toppled node will cause one neighboring node to topple. For one such toppling three classes of nodes can be defined: the unstable node is the ``active" node; the node from which it received the load is the ``parent" node; and the rest of the active node's neighbors are ``children" nodes.

In the instant before an active node topples, the parent node will have already toppled and will be empty, and the active node will have $d$ units of load and be unstable. The mean-field theory approximation allows us to assume that the children are randomly sampled from the network where the probability that a child has $i$ units of load is $\rho_i$. In the instant just after the active node topples, the parent will have 1 unit of load returned to it, and the active node will be empty, while each of the children will have $i+1$ units of load with probability $\rho_i$.

The changes to the load distribution variables $\rho_i$ caused by the toppling of the active node can be summarized as:

\begin{equation} \label{MFT}
	\begin{split}
		&N\partial_t\rho_0 = -(d-1)\rho_0 \\
		&N\partial_t\rho_1 = (d-1)(\rho_0-\rho_1)+1\\
		&N\partial_t\rho_j = (d-1)(\rho_{j-1}-\rho_j)   \qquad\qquad    2\le j\le (d-1)\\
		&N\partial_t\rho_d = (d-1)\rho_{d-1}-1
	\end{split}
\end{equation}
The first line, for example, gives the rate of change of the total number of empty nodes. Before the toppling, the parent is empty, and each of the $(d-1)$ children are empty with probability $\rho_0$. On average, this gives $1+(d-1)\rho_0$ empty nodes in the considered region. After the toppling, there is only $1$ empty node - the active node. Therefore the total number of empty nodes, $N\rho_0$, changes at the rate  $-(d-1)\rho_0$.

During the stationary state $\partial_t\rho_i=0$ for all $i$. Using this to solve Eq. \eqref{MFT} gives $\rho_0=0=\rho_d$ and $\rho_i=\tfrac{1}{d-1}$ for $1\le i\le d-1$. Then the average load per node is:

\begin{equation} \label{<s> for c=d-1}
	\langle s\rangle = \sum_{s=0}^{d-1} s \rho_s = \sum_{s=1}^{d-1} \frac{s}{d-1} = \frac{d}{2}.
\end{equation}

\begin{figure}[t]\centering
	\includegraphics[width=.7\linewidth]{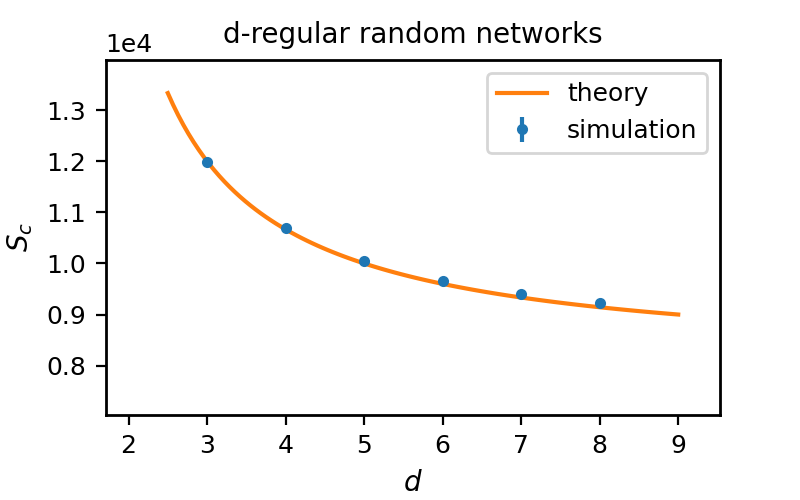}
	\caption{The stationary state load, $S_c$, of the BTW sandpile model as a function of the degree of the random regular network.}
	\label{S_vs_d}
\end{figure}

The next step is to extrapolate this result to general capacities. It is natural to expect a linear relationship between the average load per node during the stationary state and the capacity of the nodes $\langle s\rangle\propto \lfloor c \rfloor$. We take the floor function $\lfloor c \rfloor$ because only the integer part of $c$ matters in the BTW cascades (load comes in integer values). The proportionality coefficient can depend on $d$ and is denoted $\mu (d)$. Thus
\begin{equation} \label{<s>_preliminary}
	\langle s\rangle = \mu(d)\lfloor c \rfloor.
\end{equation}
The general expression \eqref{<s>_preliminary} must agree with Eq. \eqref{<s> for c=d-1} for the case $c=(d-1)$. This fixes $\mu = \tfrac{d}{2(d-1)}$. Thus, the expression of the total critical load including node capacity is:

\begin{equation} \label{S}
	S_c = N \langle s\rangle  =  \frac{\lfloor c \rfloor N d}{2(d-1)}.
\end{equation}
Here $N$ is the size of the $d$-regular random network, and $c$ is the capacity of nodes. It is expected that large networks with large capacities would hold more load, but there is also an unexpected dependence on the degree of the network: the larger the degree, the smaller the total capacity. Figure \ref{S_vs_d} shows the comparison of Eq. \eqref{S} with numerical simulations. Although the BTW sandpile model has been studied extensively, Eq. \eqref{S} and the impact of node degree is not in the literature to the best of our knowledge.

A more intuitive explanation of why the total load $S_c$ decreases with increasing network degree $d$ is as follows. In a critical state, one toppling leads to on average one more toppling among the children nodes. So, each of the $d-1$ children will topple with the probability $\frac{1}{d-1}$. On the other hand, for a child to topple two things must be true. It must receive one of the shedding units of load from the parent (which happens with probability $\frac{\text{shed load}}{\text{\# of neighbors}}=\frac{c+1}{d}$), and it must be at capacity (with probability $\rho_c$). Thus, the probability of a child to topple is $\frac{c+1}{d}\rho_c$. In the critical state this means $\frac{c+1}{d}\rho_c=\frac{1}{d-1}$, so the fraction of nodes that are at capacity is $\rho_c\propto\frac{d}{d-1}$. Since $\frac{d}{d-1}$ decreases asymptotically to 1, for higher $d$ fewer nodes will be at capacity, and the average load per node will be smaller, causing smaller critical load $S_c$ as shown in Fig. \ref{S_vs_d}.

\subsection{Location of the tipping point and frequency of the emergent cycle}\label{DK frequency section}
To calculate the period of the emergent BTW-KM cycle, we need to find the total load on the system just before a DK event starts and the total load just after it ends. Their difference indicates how much load needs to arrive between DK events, and we can calculate how long it will take to replenish that load in the build-up phase. Since the build-up phase is by orders of magnitude the longest phase, its duration is an approximation of the period of the BTW-KM cycle.

In the spirit of MFT we now assume a mean-field coupling 
\begin{equation}\label{cap_global}
	c_i = c^0 - \alpha (l-r)
\end{equation}
where the local synchronization $r_i$ is replaced by the global synchronization $r$. During the build-up phase when the network is synchronized $r=1$ there are no endogenous seeds and the system behaves as a BTW model with exogenous cascades only. The DK mechanism gets triggered only once exogenous cascades become large enough, i.e., the corresponding BTW sandpile model reaches criticality. Similarly, once a DK is triggered and the capacities are reduced, the activity persists for as long as the corresponding BTW model (with now reduced capacity) is supercritical. The capacities just before a DK are $c=c^0-\alpha(l-1)$ where $\alpha$ is the desynchronization penalty and $l$ is a parameter close to $1$. The total load just before the DK event (i.e., the load at the tipping point) is expressed using Eq. \eqref{S} as:
\begin{equation}\label{S_tipping}
	S_\text{TP} = \frac{\lfloor c^0 - \alpha(l-1) \rfloor Nd}{2(d-1)}.
\end{equation}
Given that the tipping point occurs at the end of the build-up phase when the system is fully synchronized, $r_{TP}=1$, we can express the location of the tipping point in the $(S,r)$ plane:

\begin{equation}\label{Sr_tipping}
	(S_\text{TP},r_\text{TP}) = \left(\frac{\lfloor c^0 - \alpha(l-1) \rfloor Nd}{2(d-1)}, 1\right).
\end{equation}

Just after the DK event ends, the KM phases are fully random. Then by plugging $r=0$ we get $c=c^0-\alpha l$. The total load just after the DK is approximated by:
\begin{equation}\label{S_min}
	S_0 = \frac{\lfloor c^0-\alpha l \rfloor Nd}{2(d-1)}.
\end{equation}

The load that needs to be replenished during the build-up phase is: $\Delta S= S_\text{TP}- S_0=N d\frac{\lfloor \alpha(1-l)\rfloor+\lceil \alpha l\rceil}{2(d-1)}$.
The cascades during the build-up phase are small, and almost no load is dissipated. Therefore we can assume a steady rate of load inflow $\frac{1}{\Delta T}$. Finally, this gives the frequency of a BTW-KM cycle:

\begin{equation}\label{MFT_f}
	f = \frac{2 (d-1)}{(\lfloor \alpha(1-l)\rfloor+\lceil \alpha l\rceil)N d\Delta T}.
\end{equation}
The frequency is not a trivial consequence of the desynchronization penalty $\alpha$. Instead, $f$ increases with decreasing $N$, decreasing $\alpha$, and increasing $d$.

\begin{figure}[t]\centering
	\includegraphics[width=.7\linewidth]{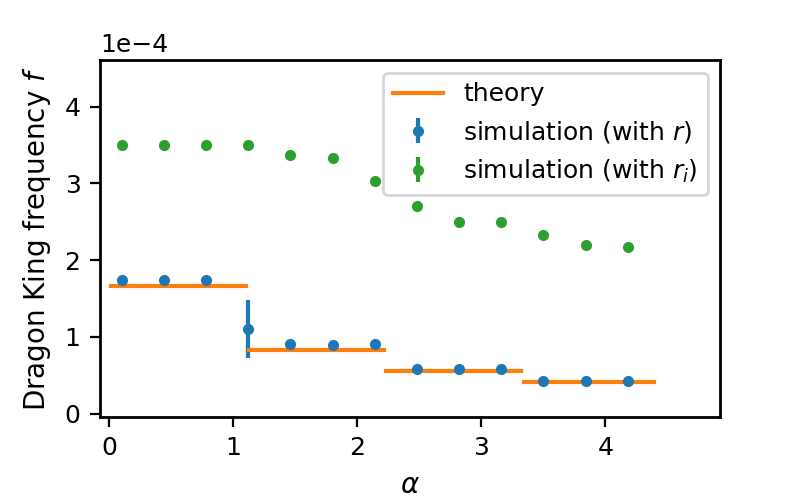}
	\caption{The frequency of the emergent BTW-KM cycle as a function of the desynchronization penalty $\alpha$. Orange is the prediction of Eq. \eqref{MFT_f}; Blue is simulation with capacities computed using global coupling Eq. \eqref{cap_global}; Green is simulation with capacities computed using local coupling per rule 2 in Sec. \ref{rules_sec}. Local coupling enhances the onset of DK events by creating more endogenous seeds and DK events also dissipate less load with local coupling, causing a higher frequency then the globally coupled MFT Eq. \eqref{MFT_f} predicts.}
	\label{model1 freq}
\end{figure}

The key equations developed under the mean-field theory assumptions (Eqs. \eqref{Sr_tipping} and \eqref{MFT_f}) provide a correct qualitative description of the model. But the approximation of the local coupling (step 2 in Sec. \ref{rules_sec}) by the global coupling Eq. \eqref{cap_global} introduces a source of considerable quantitative error (see green dots in Fig. \ref{model1 freq}). The validity of our argument however can be tested by implementing the simulations directly using MFT coupling Eq. \eqref{cap_global} (see blue dots in Fig. \ref{model1 freq}). Note, the tipping point for the load under local coupling is less then that under global coupling since the endogenous seeds start to appear and trigger a DK before the criticality is reached in this case. Similarly, the frequency for the locally coupled model is higher than the frequency in case of global coupling. The capacities are less penalized in the locally coupled model since not all $r_i$ are simultaneously zero. As a result, DKs in a locally coupled system start with less and end with more total load than for a globally coupled system. So, less load gets dissipated, and less time is required to replenish it, resulting in a higher frequency.

It is worth pointing out the applicability of our results to the cases $\alpha\to 0$ and $\alpha=0$. The presence of the ceiling function in Eq. \eqref{MFT_f} implies that the frequency will have some nonzero constant value in the limit $\alpha\to 0$. This should be expected since no matter how small the desynchronization penalty $\alpha>0$, the out-of-synch nodes will effectively have capacities decremented by $1$. On the other hand, when $\alpha=0$, the BTW-KM reduces to BTW sandpile model which has no periodicity. This is reflected in Eq. \eqref{MFT_f} which diverges for $\alpha=0$, indicating absence of periodic behavior.

\subsection{Cascade area distribution in the build-up phase}
Following the examples in the literature \cite{goh2003SFsandpile,dobson2004branching, noel2013controlling}, we construct the branching process to calculate the cascade area distribution during the long build-up phase of the BTW-KM cycle. The oscillators are fully synchronized in this phase. Therefore the capacity of each node is $c^0$. The calculation given in this section uses generating functions specific to 3-regular random networks with capacities $c^0=2$, although they can be extended to other degrees $d$ and capacities $c^0$.

First, let us consider the BTW-KM in the build-up phase with a given total load $S$, and let $p_S$ denote the probability of a random node to be at capacity. The generating function $G_S(x)$ for the cascade area can be written as:

\begin{equation}\label{branching_process}
	G_S(x) = \left(1-p_S + p_SxF_S^3(x) \right)  G_{S-1}(F_S(x))
\end{equation}
where $F_S$ is a generating function for receiving a unit of load from a neighbor:
\begin{equation}\label{branching_process_F}
	F_S(x) = 1-p_S + p_SxF_S^2(x).
\end{equation}

Each power of $x$ corresponds to a single toppled node. The total area of a cascade will be the sum of the parts started by endogenous and by exogenous seeds. Therefore the distribution of the cascade area will be the convolution of the distributions of each part, and the generating function $G_S$ is the product of the two respective parts. The first term $\left(1-p_S + p_SxF_S^3(x) \right)$ considers what happens when external load lands on a random node: it either does not topple (with probability $1-p_S$), or it topples (with probability $p_S$), thereby introducing one power of $x$, and sending one unit of load to each of its three neighbors ($F_S^3(x)$). The function $F_S$ considers the situation when load arrives on a node from a neighbor. That node, in turn, either does not topple ($1-p_S$) (the first term in Eq. \eqref{branching_process_F}) or topples ($p_S$), introducing an $x$ and shedding a unit of load to two new nodes ($F_S^2(x)$) (the second term in Eq. \eqref{branching_process_F}). The second multiplier $G_{S-1}(F_S(x))$ in Eq. \eqref{branching_process} considers the endogenous seeds. It is a fair approximation that nodes that toppled during the last cascade will become endogenous seeds with probability $p_S$. Assuming no dissipation (since we are in the build-up phase where it is rare), the previous cascade had the load $S-1$, and the number of nodes toppled during the last cascade was generated by $G_{S-1}$. These nodes will topple by probability $p_S$, sending one unit of load to two neighbors. As discussed above, this dynamics is considered by $F_S$. Therefore the area of the endogenous part of the cascade is generated by  $G_{S-1}(F_S(x))$. 

We also need to know the probability that a randomly chosen node will be at capacity which can be approximated by 
\begin{equation}\label{p_S}
	p_S=\rho_{20}+\rho_{10}(S-S_{min})/N+\rho_{00}(S-S_{min})^2/N^2.
\end{equation}
(Note this is for the BTW-KM dynamics, whereas Sec. \ref{mft SP} considered the capacities for pure BTW sandpile model.)
The variables $\rho_{i0}$ are the load distributions at the beginning of the build-up phase. $S_{min}$ and $S_{max}$ (which will be used in Eq. \eqref{G_avg}) are the lower and upper bounds on total load $S$ during the build-up phase, respectively. Equation \eqref{p_S} is the second-order polynomial approximation where the three terms can be interpreted respectively as the probability that a node was at capacity at the beginning of the build-up phase, the probability that it had one unit of load and received one more later putting it at capacity, and the probability that it was empty and received two units of load putting it at capacity. We measure the values of $\rho_{i0}$ in simulation: $\rho_{00}=0.225, \rho_{10}=0.624, \rho_{20}=0.151$.

Since $p_S$ changes as subsequent load arrives, the cascade area distribution differs at each step. To obtain the distribution of all the cascades during the build-up regime, we average over the allowed values of $S$

\begin{equation}\label{G_avg}
	G(x) = \frac{1}{S_\text{max}-S_\text{min}}\sum_{S=S_\text{min}}^{S_\text{max}}G_S(x).
\end{equation}

\begin{figure}[t]\centering
	\includegraphics[width=.7\linewidth]{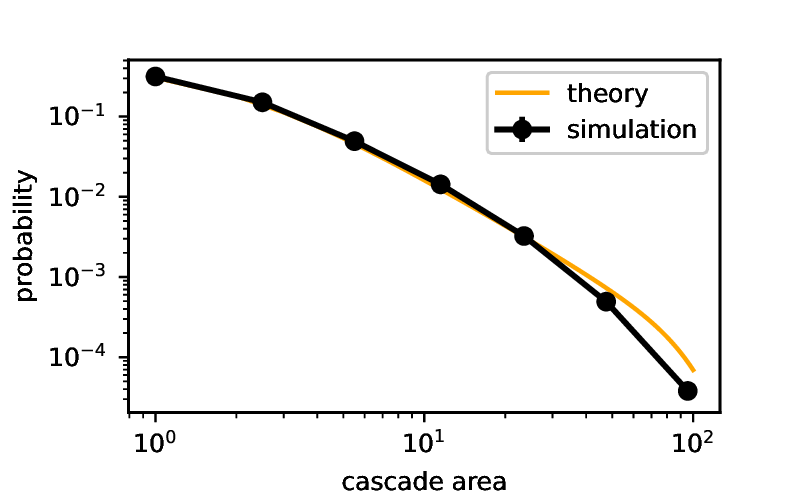}
	\caption{Cascade area distribution in the build-up phase of the BTW-KM. }
	\label{buildup_distrib}
\end{figure}

The cutoffs $S_{min}=7400$ and $S_{max}=9800$ were chosen based on analysis from the simulation to ensure that all the cascades with $S_\text{min}<S<S_\text{max}$ belong to the build-up regime. The numerical evaluation of Eq. \eqref{G_avg} is compared with simulations in Fig. \ref{buildup_distrib}. The calculation above can be generalized to other capacities and network degrees by recalculating the probabilities and accounting for the new number of neighbors and outflowing load in the branching process.

\subsection{Thermodynamic limit}\label{sec_thermo}
Finally, we discuss the thermodynamic limit of the system. Eq. \eqref{MFT_f} suggests that the frequency of the BTW-KM cycle vanishes with increasing network size. The reason is that each DK event dissipates the amount of load $\Delta S\propto N$. As $N\rightarrow\infty$ the time required to replenish the dissipated load diverges. 

\begin{figure}[t]\centering
	\includegraphics[width=.7\linewidth]{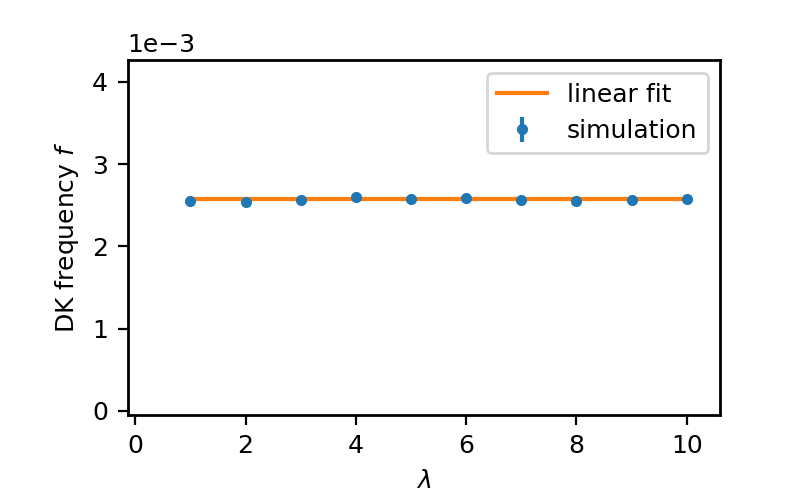}
	\caption{Emergent BTW-KM cycle frequency as a function of the parameter $\lambda$ in Eq. \eqref{Thermodynamic_limit}. Simulation results are for $N_0=1000$, $\Delta T_0=1$, and $k_0=1$ with local coupling. Dissipation is fixed at $p=0.0025$. The orange curve is a linear fit. The MFT value of frequency as predicted by Eq. \eqref{MFT_f}, likewise does not depend on $\lambda$ and is $f=1.3\times10^{-3}$.}
	\label{thermo}
\end{figure}

To avoid this divergence, we can consider the thermodynamic limit $N\rightarrow\infty$ in combination with the time scaling (see Sec. \ref{symmetries}). In other words, we look at larger systems at longer timescales. The combined transformation is

\begin{equation}\label{Thermodynamic_limit}
	\begin{cases}
		N = \lambda N_0\\
		t = \frac{t_0}{\lambda}\\
		\Delta T = \frac{\Delta T_o}{\lambda}\\
		k = \lambda k_0,
	\end{cases}
\end{equation}
and we take the limit $\lambda\rightarrow\infty$. The mean-field theoretic frequency Eq. \eqref{MFT_f} is constant under this limit.
Notice that the last three equations in \eqref{Thermodynamic_limit} are a symmetry transformation of the system. They introduce no real change in the dynamics but simply change the units that we are working with. Note that the dissipation is fixed during this scaling. Traditionally the thermodynamic limit of sandpile models with dissipation is obtained by taking two subsequent limits: network size $N\rightarrow\infty$, and then dissipation probability $p\rightarrow 0$. Therefore, for our coupled system one would have to send $p\rightarrow 0$ after first taking the $\lambda\rightarrow\infty$ limit. Figure \ref{thermo} shows  how the frequency of Dragon Kings changes in simulations under this scaling of time and network size. The simulation is performed for local coupling Eq. \eqref{cap_global}, yet the frequency is independent of $\lambda$, which suggests that mean-field prediction about correct thermodynamic scaling also holds up in case of local coupling.

\section{Discussion and conclusions}\label{sec_discussion}
Many real-world systems involve cascading dynamics on oscillator networks. We studied load-shedding cascades on a network of oscillators by coupling together the Kuramoto model and the BTW sandpile model. The introduced coupling is loosely motivated by power grid dynamics and amounts to randomly resetting node phases after toppling and making out-of-synch nodes more vulnerable to failure by lowering their load handling capacity. The system reveals an emergent long timescale and a novel phenomenon that is not present in either model in isolation: periodic cycles of Dragon King events. A DK event starts when the total load in the system reaches a tipping point when a large cascade creates endogenous seeds which start exponentially larger cascades. So the system enters into a self-amplifying feedback loop, a cascade of cascades, until it fully desynchronizes and sheds significant load.

Recall, DK events are the massive events which occur significantly more often than what extrapolation from smaller events would suggest. Often, Dragon Kings are related to tipping points, phase transitions, or bifurcations and are generated by a self-amplifying, nonlinear, endogenous mechanism that is different from the mechanism behind smaller events \cite{sornette2009dragon_kings, sornette2012dragon}. DKs are also expected to be much more predictable than black swans \cite{sornette2009dragon_kings,sornette2012dragon}. These features are all observed during the DK events of our coupled BTW-Kuramoto model.

Several aspects of the dynamics were computed analytically under the mean-field theory assumptions. We found the location of the tipping point Eq. \eqref{Sr_tipping} where the DK events kick off. We calculated the frequency of the emergent BTW-KM cycle, Eq. \eqref{MFT_f}, and Fig. \ref{model1 freq}, and showed that it depends on the network size, degree, and the strength of the coupling between the two subsystems. We demonstrate that this cycle survives in the thermodynamic limit when considered together with the time scaling symmetry transformation Eq. \eqref{Thermodynamic_limit}. Another important aspect of the dynamics is the probability distribution of the cascade area. Ideally, large cascades should be avoided, and the system should handle as much load as possible while staying synchronized. The build-up phase of the BTW-KM cycle achieves this with the corresponding cascade area distribution given in Eq. \eqref{G_avg} and Fig. \ref{buildup_distrib}. This suggests that research into control interventions to keep the system in the build-up phase could be a promising direction. We also found that the total capacity of both our coupled system and the pure BTW sandpile model is larger for lower degree random regular networks.

Similar to the BTW-Kuramoto model introduced here, there are other models with self-organization mechanisms (i.e., slow driving and dissipation) that also reveal an unexpected abundance of large events \cite{di2016self,lin2018sodk}. A crucial aspect present in each of these systems is a self-amplification of event sizes, which we explicitly quantified for BTW-KM. As a complementary approach, Di Santo et al. used mean field equations to abstract the details of dynamics away and showed that when the self-organization mechanism brings a system to an underlying first order phase transition, DK events will appear \cite{di2016self}. Our preliminary results indicate that the BTW-KM seems to belong to this class. Formalizing this would require careful further study and is beyond the scope of this present work.

The BTW-KM has many parameters. The BTW sandpile model introduces two ($c^0, p$). The Kuramoto model introduces two ($\omega, k$). And three have been introduced by us to define the coupling between them ($\alpha, \Delta T, l$). In addition, there is network size $N$ and node degree $d$ for a total of $9$ parameters ($c^0, p, \omega, k, \alpha, \Delta T, l, N, d$). Out of the nine parameters, two can be fixed without loss of generality using symmetry transformations ($\Delta T, \omega$). The emergent cycle and the DK events do not require fine-tuning of any of these parameters, but neither are they present throughout the whole parameter space. For example, at one extreme, when the desynchronization penalty vanishes $\alpha=0$, the cascading dynamics decouples from phase dynamics and the BTW model of cascading behavior is restored. Similarly, if Kuramoto coupling $k$ is large enough (threshold depending on $l$), no endogenous seeds will be possible due to quickly restored synchrony, the self-amplification mechanism will be effectively ``turned off", and critical cascading behavior of the BTW model restored. On the other hand, if dissipation probability $p$ is too large, the system will never reach the tipping point, getting stuck at a stationary fixed point that appears through an infinite-period bifurcation in the build-up phase, where average dissipation  will balance load inflow. A closer examination of different parameter regimes of BTW-KM dynamics may yield further interesting results.

There is a wealth of new phenomena in dynamical systems to explore, unlocked by coupling oscillatory and cascading dynamics. Such interactions can be implemented in various ways, of which we have studied only one natural choice. Other interesting ways of coupling the BTW and Kuramoto models include: Considering the oscillator frequency to be a function of load (to capture the slowing down of turbine generators with inertia when the demand is high); Considering Kuramoto oscillators to live on the nodes and BTW sites to live on the edges of a network (to capture the failure of transmission lines during cascading power outages); Considering two-layer network as herein, but now with each layer having a distinct topology (motivated by dynamics in visual cortex). We expect very different phenomenology from these other coupling choices. Preliminary results for the model where frequency depends on load (without phase randomization upon failure) does not show cyclic behavior or DKs. Instead, large events are suppressed, suggesting this is an interesting avenue for future investigation.

Finally, it has been proposed that DKs are more likely in homogeneous systems with strong coupling \cite{sornette1994statistical}. Here we studied BTW-KM on 3-regular random graphs. This conjecture could be explored by implementing BTW-KM on networks with broad-scale degree distributions.

\bibliographystyle{unsrt}
\bibliography{oscillator_cascades}

\end{document}